# The evolutionary approach to understand human low fertility phenomenon

Jozef Černák [1,*]

[1]*Institute of Physics, Department of Nuclear and Sub-Nuclear Physics, Faculty of Science, Pavol Jozef Šafárik University in Košice, Košice, Slovak Republic*

Correspondence*:
Corresponding Author
jozef.cernak@upjs.sk

**ABSTRACT**

Is it possible to reverse the low total fertility rate (TFR) in the developed world? Using a hypothetical model of population we have analysed the decline of the TFR which have took place in the background of ongoing global economic changes, and a liberalization process after the end of the Cold War. These phenomena have affected more that 110 millions of inhabitants of Central Europe and the Baltics and approximately 80 millions of inhabitants in Germany. The model has features of complex and evolving system of interacting individuals, and it enables to investigate a broad spectrum of input factors on individual decisions to limit the offspring. In the case of the TFR < 1.5, our initial analysis show a need of radical changes of the global economy that will stimulate series of self-regulations of demographic processes and evolution toward the safe TFR > 2.1. The changes should stimulate more uniform spatial distribution of wealth, capital and usage. They will increase a number of self-sufficient and cooperative territories, to decrease the income inequality, to decrease labour and social mobilities. Societies should investigate the impacts of economic regulations and actions on the TFR trends in advance and take into account a biological nature of women more responsible.

## 1 INTRODUCTION

In accordance with Darwin's view on World overpopulation [1], a long-term world population growth is a consequence of suitable life conditions while natural limits preventing the future growth as a lack of food or energy were not reached. In the past, serious concerns were raised about the abrupt population growth that has led to global actions increasing food safety and limiting descendants. At that time, some scientists [1] warned about ignoring biological nature of fertility and population growth, and accepting short-term solutions which would cause serious issues to the next generations [2]. However, these actions are still continuing until now, for example United Nations in New York agreed-upon Sustainable Development Goals with specific targets that in the effects will decrease the world population[2]. Population projections are notoriously uncertain [1, 3], and there are predictions that population will continue to grow [4]. The data show a slow down of its growth even some of the authors [5] claim that the world population will have been stabilized before the end of the century. Globally, the TFR has been falling over a long period, from $5.061$ in $1964$ to $2.453$ in $2014$, and has shown considerable geographical differences [6]. The global TFR slowly approaches to the critical fertility $2.1$ births per woman that is needed for a safe population replacement. In developed areas of Europe [7, 8], USA [9], and East Asia [10], fertility has been under the critical value for a long time which is considered to be a significant signature of the demographic transition [3, 7, 11]. In Africa, fertility also declines but is still above the critical value. We think that these big geographic differences of the TFR are arising from disregard of self regulation principles of population growth [1] in different environments, for example economy.

Energy crisis in $1970$ and Early recession in $1990$ are important economic and historical milestones to better understand influences of economy on declining of the TFR. In the Eastern bloc before the fall of the Iron Curtain, economic systems were characterized by central planning of both production and usage and by advanced social systems with super-welfare [7] After 1989, a transition to free market economy and governmental guarantees have declined. Later, an integration process to EU and much more higher influences of globalization have started. Very quickly after transformation, new long-term phenomena have emerged like unemployment, corruption, crime, low quality education and health care, weak pension system, increasing of taxes and living expenses, and property differentiation of individuals.

Initially, the causes of low TFR have been linked with unspecified modernization [3]. The recent theories has accepted the influences of rapid expansion of social liberalism and economic deregulation [8].

Analytical [12] and probabilistic [2, 13] models of population have limited possibilities to link economic parameters into demographics simulations directly. These models are based only on demographic parameters where economic influences are taken into account indirectly. They are able to consider gender roles [12, 13], however the details about men's behaviour in globalized society are not well know due to lack of information [14, 15]. We assume that advanced computational models based on learning and evolving agents should remove these drawbacks of analytical methods. Evolutionary scenarios could provide additional information about impact of globalization on social interactions between women and men.

Our ambitions are going behind the current trends focused on projections at the current economy [2, 4, 5, 13]. In the future, we plan to investigate theoretical and computational models that will stimulate self-regulation of population closer to the secure TFR. To do this we consider human population for a living system that interacts with environment while obeying evolutionary rules [16].

## 2 METHOD AND DATA

The TFR and population statistics used in this study are available in the World Bank [6]. We have proposed a hypothetical computational model (see Sec. 3.1) of population as a complex system [17] with features to learn and evolve. It was used to identify the most possible economic scenaria leading to low TFR $< 1.5$. These case studies are useful to understand global properties of population as well as individual decisions. We have analysed thought experiments and proposed hypothesis. However, they should be tested by other computational and mathematical methods.

The fall of the Berlin wall in $1989$ and Early recession in $1990$ have had a strong effect on demography in Central Europe and the Baltics. In Fig. 1 a deep declining of the TFR is shown where the TFR has remained under the EU average until now. In the consequence of two simultaneous dominant effects, i.e. the long-term TFR below the replacement level and free labour mobility supported by EU the population continuously decreases (Fig. 1). The TFR statistics of Germany and the former USSR country, Kazakhstan, are used to demonstrate different demographic trends. In Germany the transition to the the log-term low TFR < 1.5 is coupled with Prague spring in 1968 and Energy crisis in 1970. In Kazakhstan the TFR declined after the end of the Cold War. However, later in 1999 the TFR has been reversed and now is above the safe TFR level.

## 3 SIMPLIFIED MODEL OF POPULATION

Sociologists are looking for solution-oriented methods to make progress in social science [18]. We consider a treatment of the low TFR for such task where application of mathematical, physical and computational methods will lead to progress in understanding of this phenomenon. Quantitative analyses of social systems are difficult due to obstacles to find a single theory and appropriate parameters to describe the investigated phenomenon[19]. This is a reason why we present initial ideas instead of the quantitative results.

### 3.1 Model specifications

To understand an individual choice of a woman to reduce offspring as well as influence of economic factors on the TFR we consider a simplified model of human population as a complex system [17, 20, 21, 22] with capabilities to learn and evolve [16]. The model consists of many F and M individuals that interact with one another and with environment through diverse interactions for example, economic [23] and social [24]. Multiplication of individuals, F and M, mimics a sexual connection, and F could breed the next generation. Similarly, their instincts mimic basic human instincts and they can learn and evolve. Let us consider that individuals belong to multi-layered economic[23, 25], social [26], cognitive [27] and other networks [28], then they are considered to be nodes of these networks. Each network has own topology, development rules and properties of links. Each link has assigned a variable that is proportional to the strength of the interaction between nodes. Nodes can move in the space and create different space arrangements, i.e. aggregates, networks and etc. see Fig. 2. The strength of interactions among nodes can evolve depending on unspecified factors like distance between nodes, or link history. Each change of the system involves huge series of the responses of individuals due to mutual interactions and positive and negative feedbacks that are typical for complex systems while some of responses are hidden before individuals. Transitions to new states are reached by adaptation, learning, self-organization and collective phenomena [17, 22]. The individual response of F (M) on surrounding stimulus i.e the choice could be simplified to an artificial neuron where its input signals (inputs) are assigned to each network link. The inputs are proportional to the excitations and in the node they are weighted and used as arguments of a node function. If value of the node function is higher than a threshold, a binary output is changed, an answer is: yes/no. Weights and the node function are unique for each node and reflect a state of F and M in time of the choice where learning of F (M) changes the node function. All parameters of the model can evolve. In the next section we assume that economic, social and information networks are needed to cover broad spectrum of population behaviours. We propose three case studies that shed light on the individual choice of F(M) that is influenced with economic factors.

### 3.2 Thought experiments

The individual choice of F (M) is influenced by many diverse inputs (links with surrounding nodes) as well as the node function. Some inputs are coupled and they act on F (M) simultaneously. Let us consider only two limited cases of influences of surroundings on individuals. In the case of weak interactions with environment, low levels of input excitations and low weights, the choice depends only on individual behaviour of F, i.e. her node function. In the case of strong interaction with environment, some inputs are strong and weights are high, a choice of F is influenced by a few or many external factors. The fact that many of F individuals in population choose the same answer, i.e. to limit descendants, needs serious attention. We have proposed three thought experiments to better understand universal experimental observation that if living standard increases the TFR decreases [1].

### 3.2.1 Low economic safety of F and M

In developed world, most of F (M) individuals work and have high living standards. A detailed view shows that their prosperity is really very unstable and depends on many factors like a permanent income, personal savings, loans, wealth, capital etc. [23, 29] If individuals lose their job and remain unemployed for a long time their living standard gradually decreases due to low savings and after a certain time they are socially isolated. Most of F (M) individuals evaluate this situation as a serious threat to their survival.

### 3.2.2 Equality of F and M

A society promotes an idea of equality between F and M strongly and does not recognize biological differences between them when they compete for the same working position. This obliges F to behave like M and live without descendants. It is evident that a choice of F to have descendants has a negative impact on her presence in the economic network and this influences her survival (see the case above). Her choice is significantly biased.

### 3.2.3 Collective experience of F and M

Another important case could be a collective experience (illusion) of F (M) obtained by learning (changes of node functions) that in high living standards it is easier to survive without descendants.

Generally, there are many other reasons that influence F in her choice [3, 23, 24].

## 3.3 A possible impact of economic regulations on TFR-hypothesis

Core-periphery urban areas (see Fig. 2 ) are ubiquitous in nature, for example a castle – lower castle, city – suburbs, USA – Europe, North – South hemispheres. They show the common universal underlying phenomenon of their formation, i.e. a self-organization of complex systems [17] that reflects economic and power rules in societies. Generally, systems of many spatially distributed and interacting elements (population and economic markets) are characterized by many degrees of freedom and their parameters can show the diverse spectrum of behaviours that makes difficulties to control them and predict future states of the systems. Typical example is a population growth and the TFR [1, 4, 5]. In the effort to control some of the parameters in society new control mechanisms and rules are added. However, the society may respond to these changes by unwanted manner and a part or the whole system could diverge in the unwanted states, for example a slow economic growth, unemployment of young [25] and the low TFR in Germany [13], Central Europe and the Batics (Fig. 1). The TFR in China [30] is unexpected but still in the regime TFR $> 1.5$ where institutional politics [30] could increase the TFR [7, 8]. We consider long-term low TFR $< 1.5$ for such unwanted state of a complex system, i.e the state could be a consequence of economic regulations to reach sustain economic growth and low unemployment in societies.

## 4 DISCUSSION

In the demographic data shown in the Fig. 1 we have identified three important periods, first 1960-1968 second 1968-1989 and third 1989-2014. In the first period, the TFR of Western Europe was higher than in Eastern Europe. The TFR in these competing parts of Europe was above the safe replacement TFR $> 2.1$. The TFR trends changed significantly in the beginning of the second period when in Germany and Western Europe the TFR has dropped below the safe level of the TFR $< 2.1$. In Germany has fallen to the low TFR $< 1.5$ dramatically. At that time in Eastern Europe, a demographic development was quite different. The TFR has decreased only gradually and has been above the safe replacement level. Very shortly after the beginning of third period, the TRF in Eastern Europe has dropped deeply below the low TFR $< 1.5$ and it persists until now. The character of this demographic transition is similar to transition in Germany twenty years ago (Fig. 1). On the other hand, the decrease of the TFR in Eastern Europe has been incorrectly considered for a temporal phenomenon [7]. The demographers [7] have not carefully analysed important economic and political impacts on the TFR in Germany in 1968 as well as the whole Western Europe. They also have neglected Kirk's [3] observations that a low fertility phenomenon has tendency to spread quickly and to persist for a long time that we can demonstrate in the Fig. 1.

In the former Eastern Europe, at the same time the decrease of the TFR, and emerging labour mobility together with political and economic interventions of EU [25] have caused long-term declining of population as it is shown in Fig. 1. The mobility has essentially changed the age profile of the population due to higher mobility of young and educated individuals [25]. EU founding countries suffer from low fertility the whole time [3] thus a permanent inflow of qualified labour from new territories assists them to fill in missing work positions [25] even to increase population. The EU stagnation is strengthen by weak prognosis to extend EU common market into new territories [31] and EU political decisions that prefer ideological goals instead of economic laws [25]. We assume that strong political efforts of EU officials to extend and integrate EU should guarantee a safe population replacement only in the most developed parts of EU (Fig. 2a) in the long term perspective without a need to solve low fertility issue in Germany, Central Europe and the Baltics.

Demographic data shown in Fig. 1 fully confirm Stiglitz's [25] conclusions that Europe is divided on core that is large territory West-North orientation where population increases and a large territory in South-East orientation where population decreases see Figs. 1, 2a. Population projections for Eastern Europe [32, 33] are very alarming since they have agreed on the conclusion that population will continue declining in the long-term perspective. We are afraid of these population trends [5, 13] that can cause a shadow future in this region.

The long-term TFR under a replacement on large territories is a serious biological issue that can lead to extinction of many societies and decreases diversity [16]. It could also be a serious economical problem because the current global economy needs to increase global annual production that affects global usage. The increase of usage can be reached by increasing the number of consumers (a population growth) or by increasing individual usage where both solutions have real limits. One of proposed scenaria is the moderately lower fertility with modest population decline that will force individuals to consume much more and will favour higher living standards [32]. This scenario with the TFR above $1.6$ [32] is not sustainable and will lead to slow population declining [8]. We assume that increase of living standard [1] in this model will again lead to a transition to low TFR $< 1.5$.

We have analysed several historical events similarly as Goodhart [1] have reported to identify possible actions to increase TFR. For example, a shift from hunter to agrarian society slowed-down the motion of individuals as a consequence of new economy that emerged at this time. Population dramatically increased shortly after WW II (1945-1960) when the world population temporarily produced really less than it needed and economic activities took place simultaneously on large territories, i.e. production and usage were much more uniformly distributed across the globe (Fig. 2b). This led to rapid economical growth [23] without need to control it and lower inequality [29] in population. We can see similarities between economic and social conditions in Eastern Europe in 1968-1989 and France and Sweden where TFR is high. The most visible common features of these countries are low inequality [29], stable employment near home, easier way to create a new household and an institutional support for families.

McDonald [8] assumes that low fertility is unintended rather than a deliberate outcome of changing social and economic institutions. Considering many examples in ancient and recent history (Fig.1) we believe that emerging global low TFR phenomenon is a natural consequence of biological nature of fertility and population growth [1] that is influenced by the environment. The TFR in Fig. 1 shows very sensitive responses of population on political-economic changes. We found that a reverse of fertility in developed countries is a specific and rare phenomenon [34] thus replacement of population needs more attention, for example looking for new long-term and sustainable solutions to keep the TFR around the safe level. One of possible approaches is to investigate much more advanced models of population that will enable to take into account actual economic and political factors as well as social influences.

Economic prognoses predict a lower economic growth [23]. However, the low TFR has negative impact on economy [33]. We have proposed an idea that economic changes should stimulate more uniform spatial distribution of world production and wealth usage (see Fig 2b), increase personal savings and capital, increase in a number of self-sufficient and cooperative territories, decrease of labour mobility, decrease of income inequality. Careful consideration of both the economic regulations and biological nature of women fertility in social actions are needed.

At least two pillars of modern society, i.e. mobility and equality, should respect more the natural rules of evolution and should be revised. Decrease of the mobility is an important factor in order to create stronger social bonds (stronger links in social network) [26]. Advanced computer simulations [35] show that monogamy is the result of evolution, however at present we can see many opposite trends, for example new sexual behaviours of men and women [14, 15], that are a consequence of globalization and they lead to dissolution of families. Revision of equality should ensure safe conditions for motherhood and parenthood in time that is optimal for women. All these factors, i.e. economical safety of women, accepting their biological nature of fertility, strengthen influence of their closest together with legislative support for motherhood [30, 36] and parenthood could change their individual decisions to accept offspring. Generally, it might be very useful to promote a standard life style and moral values in accordance with evolution rules [16] where descendants are an integral part of population. We think that these changes could stimulate series of secondary self-regulation processes [1] and evolution toward the safe fertility.

Thought experiments Sec. 3.2 could be applied in each country to analyse impact of a selected economic cases on the individual decisions. For example, all three thought experiments could be used to analyse situation in Germany, Central Europe, the Baltics and Kazakhstan shown in Fig. 1. We assume that higher living standard causes lower economic safety of individuals Sec. 3.2.1 that could be one of the possible explanations why higher living standard causes lower fertility [1]. Similarly, equality between women and men Sec. 3.2.2 influences women's individual decisions, in addition it could cause tensions between women and men[8] when they are competing for the same working position. Education of individuals, collective experience and social actions of linked individuals Sec. 3.2.3 are also important factors to better understand the low TFR trend in Germany. We believe that understanding of demographic transition    in Germany in 1970 is important to treat the low TFR in Eastern Europe where the similar transition has taken place 20 years later (Fig. 1). We assume that main reason of the low TFR in Germany is a consequence of specific economic regulations that are focused on: currency stability, economic growth, employment levels, and trade balance. All these economic parameters have been controlled from 1967 until now [25]. It is evident that this controlling has a long-term negative consequence on the TFR trends see Fig. 1 while details are not well know but they could be uncovered by advanced computer models (Sec. 3.1).

The end of the Cold War also influenced the TFR of Kazakhstan [37] (Fig. 1). Economic and geographic isolation of Kazakhstan from EU should be one of the significant reasons that the TFR was turned. In the opposite, the former USSR countries integrated in EU, i.e. the Baltics are below the safe replacement level for a log-time with a low perspective to increase the TFR.

There are visions that social science have to adopt more solution oriented approaches [18] to solve realistic social problems. We indicate such approach suitable for developed societies to investigate economic and political impacts on the TFR trends. Demographic transition and globalization diminish importance of childbearing in women's lives, however one important question remains still opened: "What will fill this space [24]?"

# 5 CONCLUSIONS

Demographic data of more than 190 millions of European inhabitants were used to demonstrate an impact of economic and political changes during a transition from centralized to a "free market" oriented economy, more precisely to the economy with many regulations to ensure a sustainable economic growth and low unemployment. Time series of the TFR show that the TFR trends in Europe are very sensitive on integration and globalization efforts as well as economic regulations. We assume that these economic and political regulation actions cause persistent low fertility in Germany, Central and East Europe. Long term population projections are alarming for developed world, especially for Europe, with no signature to increase the low TFR in a sustainable way. We assume that mathematical and physical methods and tools should be used to treat dynamical systems in which variables show non-Gaussian statistics. We have proposed a hypothetical model of population as a complex and evolving system that enabled us to integrate economic and social models to study the individual woman's choices to limit descendants and could help to identify primary reasons of the low TFR. We believe that radical changes of global economy are needed to increase the low TFR. This will stimulate self-regulation and evolution toward the safe TFR. The changes should initiate the evolution towards more uniform spatial distribution of production of wealth and capital, more uniform distribution of usage, increase in a number of self-sufficient and collaborative territories, increase of individual savings and capital, decrease of labour mobilities, decrease of inequality and careful consideration of economic regulations and a biological nature of women fertility in economic and social actions. These proposals could be tested in advance using computational models of interacting and evolving agents to see the impacts on the TFR trends before application of "new rules" in societies.

## ACKNOWLEDGMENTS

I acknowledge my mother, Z. Eperješiová, S. Bartošová and M. Reľovská for thought-provoking discussions and V. Szappanošová for reading manuscript.

## FIGURE CAPTIONS

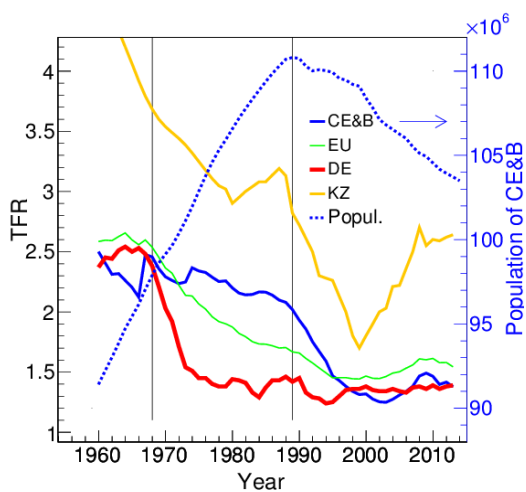

**Figure 1.** Decline of the total fertility rate (TFR) and population (Popul.). Trends in TFR (left y axis) of European Union (EU), Central Europe, and the Baltics (CE&B) and population of CE&B (right y axis). The list of CE&B countries: Bulgaria, Croatia, Czech Republic, Estonia, Hungary, Latvia, Lithuania, Poland, Romania, Slovak Republic, and Slovenia. For comparison, TFR of Germany (DE) and Kazakhstan (KZ) are shown. Kazakhstan is a former USSR country that was affected by the end of the Cold War, however shows a turn to the safe TFR. Vertical lines show important historical events, i.e. Prague Spring in 1968 and removing the Iron Curtain in 1989.

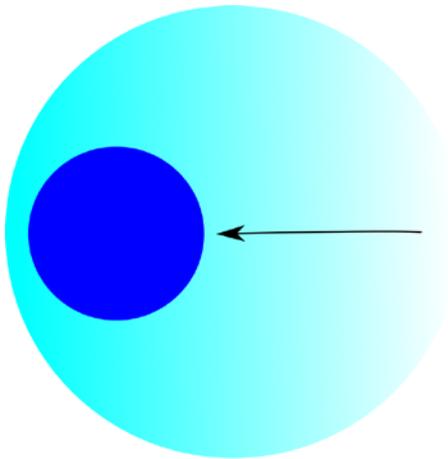 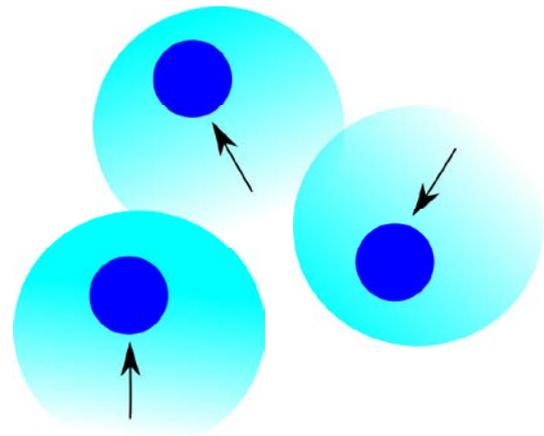

**(2a)** Core-periphery economic structure.

**(2b)** The same territory as is shown in Fig. 2a is divided into smaller self-sufficient and cooperative territories of production and usage to slow down mobility.

**Figure 2.** Spatial distribution of production and usage. Arrows show labor mobility, core (a wealth territory) is dark blue and periphery (a poor territory) is light blue.